\documentclass{PoS}
\pdfoutput=1
\usepackage{siunitx}
\usepackage{url}
\usepackage{booktabs}

\title{Searching for GRBs at VHE with MAGIC: the status before CTA}

\ShortTitle{Searching for GRBs at VHE with MAGIC: the status before CTA}

\author{\speaker{Alessio Berti}$^{1}$, Lucio Angelo Antonelli$^{2}$, Zeljka Bo\v{s}njak$^{3}$, Juan Cortina$^{4}$, Stefano Covino$^{5}$, Valerio D'Elia $^{2,6}$, Elia do Souto Espi\~neira$^{7}$, Satoshi Fukami$^{8}$, Susumu Inoue$^{9}$, Francesco Longo$^{10}$, Davide Miceli$^{11}$, Elena Moretti$^{7}$, Lara Nava$^{5}$, Koji Noda$^{8}$, Seya Nozaki$^{12}$, Michele Peresano$^{11}$, Yusuke Suda$^{13}$, Martin Will$^{13}$ for the MAGIC collaboration\footnote{\texttt{https://magic.mpp.mpg.de/}. For collaboration list see PoS(ICRC2019)1177.} \\
$^{1}$University of Torino and INFN Torino, Torino, Italy \\
$^{2}$INAF, National Institute for Astrophysics, Rome, Italy \\
$^{3}$Faculty of Electrical Engineering and Computing, University of Zagreb, Zagreb, Croatia \\
$^{4}$CIEMAT, Madrid, Spain $^{5}$INAF, Osservatorio Astronomico di Brera, Merate, Italy \\
$^{6}$Space Science Data Center - Agenzia Spaziale Italiana, Rome, Italy \\
$^{7}$IFAE-BIST, Bellaterra (Barcelona), Spain $^{8}$ICRR, University of Tokyo, Kashiwa, Chiba, Japan \\
$^{9}$RIKEN, Wako, Saitama, Japan $^{10}$University of Trieste and INFN Trieste, Trieste, Italy \\
$^{11}$University of Udine and INFN Trieste, Udine, Italy \\
$^{12}$Department of Physics, Kyoto University, Kyoto, Japan \\
$^{13}$Max-Planck-Institut f\"ur Physik, Munich, Germany \\
E-mail: \email{Alessio.Berti@to.infn.it}}

\abstract{Gamma-Ray Bursts (GRBs) are one of the main targets for current and next generation Imaging Atmospheric Cherenkov Telescopes (IACTs). Given their transient behavior, especially in the case of their prompt emission phase, performing fast follow-up observations is challenging for IACTs, which have a narrow field of view and limited duty cycle. Despite this, MAGIC plays a major role in the search for Very High Energy (VHE, E>100 GeV) gamma-ray emission from GRBs: this is possible thanks to its fast repositioning speed, low energy threshold and high sensitivity at the lowest energies. In 2013 the MAGIC GRB automatic procedure was upgraded, increasing the number of GRBs followed in the prompt and early afterglow phases and decreasing dramatically hardware failures during fast repositioning. Currently, only GRB\,190114C was firmly detected in the VHE band, while for other GRBs no significant detection was achieved. In such a case, upper limits (ULs) can give insight into the physics driving such eluding sources, especially on their emission mechanisms. In this contribution we report on the status of the GRB follow-up with MAGIC and focus on the ULs and results obtained from a sample of GRBs observed between 2013 and 2018. This GRB catalog is the result of the MAGIC well-designed and tested follow-up procedure, and it serves as a precursor of GRBs observation with the next generation IACT system, the Cherenkov Telescope Array (CTA).}

\FullConference{36th International Cosmic Ray Conference -ICRC2019-\\
		July 24th - August 1st, 2019\\
		Madison, WI, U.S.A.}

\begin{document}

\section{Introduction}
Gamma-Ray Bursts (GRBs) are one of the most elusive and enigmatic class of sources. They are transient sources exhibiting an explosive behavior, causing a huge release of energy in a small amount of time. One of the main questions about GRBs concerns their emission of gamma-rays at very high energies (VHE, $E\gtrsim\SI{100}{\GeV}$). GRBs have been observed to emit photons in a wide range of the electromagnetic spectrum, from radio waves to high energy (HE, $E\lesssim\SI{100}{\GeV}$) gamma-rays and, since January 2019, also in the VHE band in the case of GRB\,190114C \cite{Mirzoyanetal2019a,Mirzoyanetal2019b}. Still, most of the information on the HE emission comes from the data collected by the \textit{Large Area Telescope} (LAT) instrument onboard the \textit{Fermi} satellite. The high energy emission is quite common and exhibits some recurring features: it is delayed with respect to the lower energy emission and it has a long lasting duration \cite{Ajelloetal2019}. Despite that, GRBs detected at GeV energies help in driving theoretical models, a common interpretation has not been achieved yet. The \si{\GeV} emission is most likely produced by synchrotron radiation in the afterglow, but in some GRBs a second component is needed to describe the spectra. The detection of a VHE signal from GRBs would help to understand better the origin of HE emission. Among others, Cherenkov telescopes contribute to the observation of GRBs in the VHE range. They operate in an energy range starting from few tens of GeV, providing a good overlap with \textit{Fermi}-LAT but with higher sensitivity. \\
In this context, among the Imaging Atmospheric Cherenkov Telescopes (IACTs) currently in operation, MAGIC plays a major role in the follow-up of GRBs at VHE. In this contribution we present the GRB follow-up program within MAGIC and the possible theoretical frameworks predicting VHE emission from GRBs. 

\section{VHE emission models for GRBs}
In first approximation, extending the low-energy processes to higher energy could be a viable explanation for the possible VHE emission from GRBs. One example is considering the synchrotron emission produced by relativistic electrons in the forward shock of the afterglow \cite{kumar_duran_10}. It is shown that for some GRBs the emission at HE can be used as a proxy to predict the optical/X-ray bands flux and viceversa. The origin of such synchrotron emission can be either leptonic \cite{zhang_meszaros_01} or hadronic \cite{bottcher_dermer_98, peer_waxman_04}. \\
A second possibility is self-synchrotron Compton (SSC) emission in the afterglow \cite{sari_esin_01}. This model however can have issues in explaining the delayed emission in the \si{\MeV}-\si{\GeV} band. In the case of MAGIC observations, such model was adopted to discuss its viability in relation to the upper limits derived on GRB\,080430 \cite{aleksicetal2010} and GRB\,090102 \cite{aleksicetal2014}. \\
Other models consider the emission of VHE photons as the result of the upscattering of prompt $\gamma$-rays by electrons \cite{meszaros_rees94, beloborodov05, fan_wei_05}, CMB photons upscattering \cite{plaga95} or Compton scattering of thermal emission with electrons in the photosphere \cite{toma_etal11}. \\
The detection of the possible VHE emission from GRBs, beside improving the knowledge of GRB physics (bulk Lorentz factor determination, magnetic field amplification, acceleration of particles in the jet), could give hints to solve some pending issues like the possible role of GRBs in the production of UHECRs or neutrinos \cite{waxman_95, pac_xu_94}.

\section{Follow-up of Gamma-Ray Bursts with MAGIC}
\subsection{The MAGIC Telescopes}
The \textit{Major Atmospheric Gamma Imaging Cherenkov} (MAGIC) system comprises two Cherenkov telescopes of \SI{17}{\meter} diameter. The MAGIC site is located within the Observatorio Roque de Los Muchachos (ORM) at \SI{2200}{\meter} a.s.l. in the Canary island of La Palma, Spain (\ang{28.8} N, \ang{17.9} W) \cite{Aleksicetal2016a}. Currently operating for more than 15 years, in a first phase MAGIC operated as a single telescope up to 2009, when a second telescope was added. Therefore MAGIC performs its observations in stereo mode. The performance of the system improved in time thanks to several hardware interventions (e.g. the major upgrade in 2011-2012) and to new analysis techniques, leading to the current capabilities of the telescopes in detecting VHE gamma-rays \cite{Aleksicetal2016b}:
\begin{itemize}
\item integral sensitivity (in \SI{50}{\hour}) of $\sim0.7\%$ of the Crab Nebula flux above \SI{220}{\GeV}
\item an angular resolution of $\sim\ang{0.06}$ at \SI{1}{\TeV} (\ang{0.1} at \SI{100}{\GeV})
\item an energy resolution of 15\% at \SI{1}{\TeV} (24\% at \SI{100}{\GeV})
\end{itemize}
MAGIC has an energy threshold of $\sim\SI{50}{\GeV}$ (at trigger level and at zenith), making it suitable to observe and possibly detect far-away sources, like most GRBs, for which the the effect of the gamma rays absorption on the Extragalactic Background Light (EBL) is relevant at few hundreds of \si{\GeV}. \\

\subsection{The MAGIC GRB follow-up strategy}
The follow-up of GRBs has always been one of the main scientific projects of the MAGIC collaboration. Beside the aforementioned figures of merit, MAGIC light structure allows to move the telescopes at high speed ($\ang{7}$/s in the so-called \textit{fast mode}). This is a crucial point for the observation of GRBs: they are transient sources which can be promptly localized only by instruments with a large field of view, like satellite-based experiment (e.g. \textit{Fermi}-GBM, Swift-BAT, AGILE). \\
MAGIC instead, like other IACTs, has a limited field of view (\ang{3.5} diameter). Therefore it relies on external triggers in order to perform a follow-up. This exchange of information is made possible thanks to the \textit{Gamma-ray Coordinate Network} (GCN, \url{https://gcn.gsfc.nasa.gov/}), which collects and distributes the alerts on GRBs and transient events (e.g. high energy neutrinos and gravitational waves) to interested partners. MAGIC receives these alerts thanks to its Automatic Alert System (AAS), which has the task to process them and evaluate the visibility of the targets from the MAGIC site, according to predefined criteria. If a GRB is observable immediately after receiving the alert and the telescopes are prepared for observations, the so-called automatic GRB procedure is triggered. Irrespective of the source being observed at the time of the alert, the telescopes are moved in fast mode towards the GRB position. During the slewing movement, other subsystems are prepared to start data taking as soon as the telescopes reach the final position: the DAQ is re-initialized, the mirrors are adjusted and focused, triggers and thresholds are set. If an alert does not meet the criteria at the moment it is received, but it will at some point in time within four hours from the time of the GRB trigger, the observers at the MAGIC site are informed, so that they can prepare for the possible observation. In any case, they will contact the Burst Advocate (BA) on duty. The role of the BA is to support the observers in the case of a GRB alert, mainly evaluating the relevance of the alert (if it is real or not) and monitoring the observation. The BA can therefore choose to stop the observation ahead of time (usually GRBs are followed up to four hours after the time of the trigger) or to prolong it in the case of hints of a signal at VHE or presence of interesting counterparts in other wavelengths and/or messengers. \\
The automatic GRB procedure was upgraded in 2013 in order to reduce at a minimum possible technical problems. In particular, the DAQ system was adapted so that it is not stopped during the movement. When reaching the final position, the telescopes will observe the target using the standard wobble mode. In most cases, GRBs are observed up to four hours from the trigger but in some cases the observation is performed even at later times. For this reason, we refer to them as ``late time observations''. The purpose of such an approach is to try to detect the possible late afterglow emission from GRBs. Therefore, the most suitable targets of late time observations are GRBs detected by \textit{Fermi}-LAT. This strategy is carried out whenever a GRB is detected by the LAT and its observability is good from the MAGIC site. If the GRB trigger comes during the day, the observation can be scheduled for the coming night. If the GRB trigger comes during the night, since the information about LAT detection has a delay of few hours, the late time observation can be scheduled for the next night.

\begin{figure}[!t]
\centering
\includegraphics[width=\linewidth]{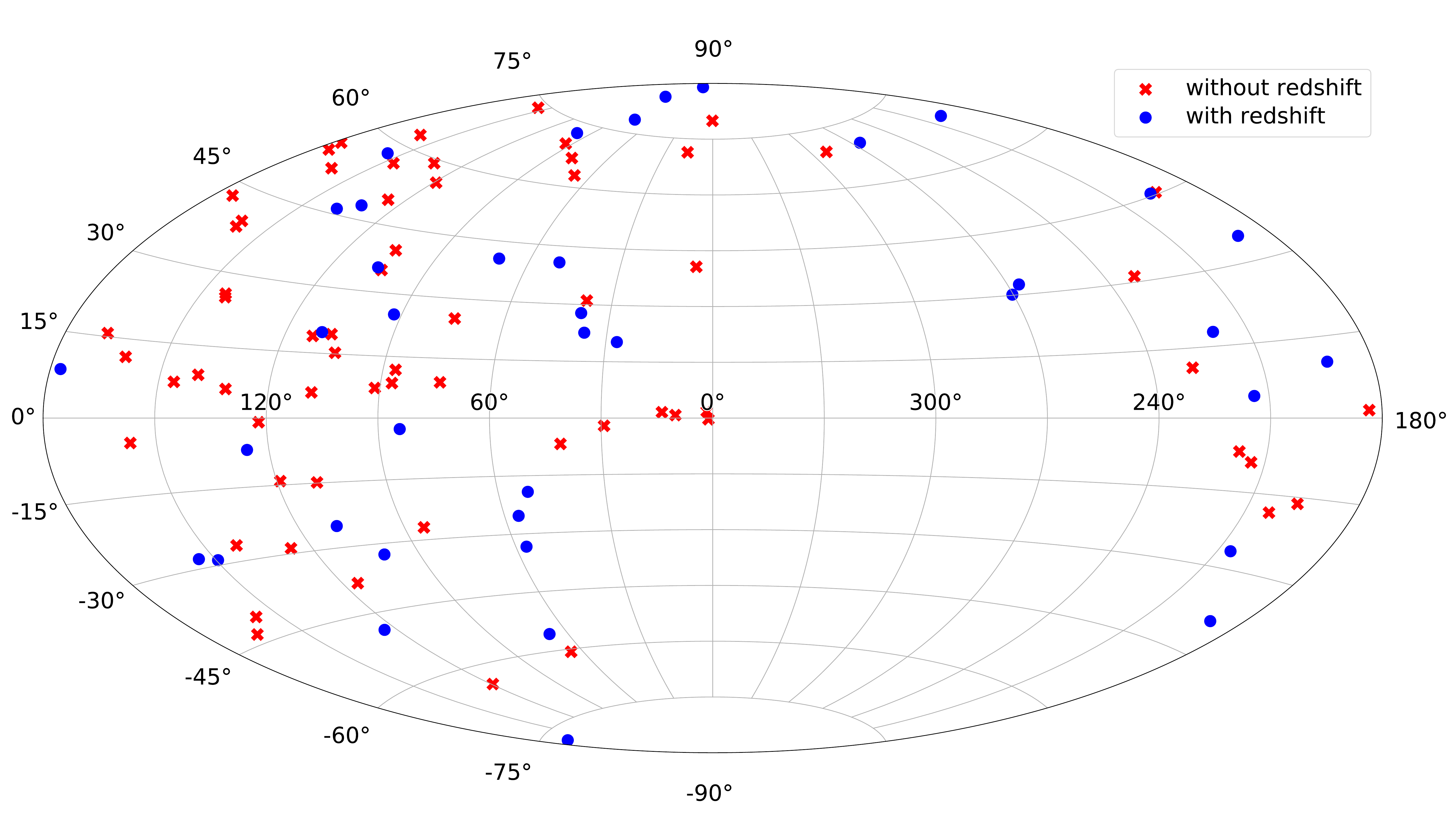}
\caption{Skymap showing the locations of the 103 GRBs followed-up by MAGIC up to December 2018. Filled blue dots are GRBs with redshift estimation, while red crosses are GRBs without redshift.}
\label{fig:magicgrbskymapuptodec18}
\end{figure}

\begin{table*}[!t]

\begin{center}
\begin{tabular}{cccccll}
\toprule
GRB name   & T$_0$ & T$_{90}$ & z & Zenith & Delay & Notes \\
   & \small{[UTC]} & \small{[s]} &  & \small{[deg]} & \small{[s]} &\\
\midrule
GRB\,150819A & 00:50:08 &  52.1    & -       & 40       &  4888     &             \\
GRB\,151118A & 03:06:30 &  23.4    & -       & 40       &  54       &             \\
GRB\,151215A & 03:01:28 &  17.8    & 2.59    & 15       &  30       &             \\
GRB\,160119A & 03:06:08 &  116     & -       & 56       &  661      &             \\
GRB\,160203A & 02:13:10 &  20.1    & 3.52    & 55       &  195      &             \\
GRB\,160310A & 00:22:58 &  18.2    & -       & 35       &  72485    &  LAT; late  \\
GRB\,160313A & 02:37:15 &  42.6    & -       & 30       &  119      &             \\
GRB\,160504A & 19:30:37 &  53.9    & -       & 28       &  5152     &             \\
GRB\,160509A & 08:59:04 &  370     & 1.17    & 50       &  217323   &  LAT; late  \\
GRB\,160623A & 05:00:34 &  50      & 0.367   & 27       &  75897    &  LAT; late  \\
GRB\,160625B & 22:43:24 &  700     & 1.4     & 60       &  2765     &  LAT; late  \\
GRB\,160821B & 22:29:13 & 0.48     & 0.16    & 34       &  24       &             \\
GRB\,160910A & 17:19:39 &  -       & -       & 45       &  10934    &  LAT        \\
GRB\,160927A & 18:04:49 &  0.48    & -       & 32       &  6939     &             \\
GRB\,161229A & 21:03:49 &  -       & -       & 26       &  7325     &             \\
GRB\,170306B & 14:07:22 &  18.7    & -       & 22       &  36636    &  LAT; late  \\
GRB\,170728B & 23:03:19 &  47.7    & -       & 46       &  35       &  LAT        \\
GRB\,170921B & 23:07:10 &  39.4    & -       & 48       &  2767     &  LAT        \\
GRB\,171020A & 11:49:15 &  41.9    & -       & 40       &  104      &             \\
GRB\,171210A & 11:49:15 &  12      & -       & 31       &  31435    &  LAT; late  \\
GRB\,180512A & 22:01:47 &  24.0    & -       & 18       &  153      &             \\
GRB\,180715A & 18:07:05 &  0.68    & -       & 29       &  12019    &             \\
GRB\,180720C & 22:23:57 &  124.2   & -       & 56       &  108      &             \\
GRB\,180904A & 21:28:32 &  5.39    & -       & 21       &  96       &             \\
GRB\,181225A & 11:44:10 &  41.5    & -       & 47.6     &  115916   &  LAT; late  \\
\bottomrule
\end{tabular}
\end{center}
\caption[Breaks Table]{Summary of the main parameters of GRBs observed by MAGIC in stereo mode between May 2015 and December 2018. The reported T$_{90}$ are in the \num{15}-\SI{150}{\keV}. The values of the zenith distance are those at the beginning of the observation.  The delay is computed as the difference, in seconds, between the trigger time of the GRB ($T_0$) and the beginning of the observation. Notes: LAT means that the GRB had a \textit{Fermi}-LAT detection. ``Late'' means that a late time observation was performed.}
\label{tab:grb} 
\end{table*}

\subsection{GRBs observed by MAGIC}
Up to December 2018, MAGIC has followed-up 103 GRBs with favorable conditions (good or decent weather and without technical problems preventing the analysis of the data), see Figure\,\ref{fig:magicgrbskymapuptodec18}. Less than half of the total number of events has a redshift estimation: among these, seven have $z<1$. This means that for most of the MAGIC GRB sample, the EBL effect can be severe. At $z=1$, taking as reference the EBL model by Dominguez et al. \cite{dominguezetal2011}, the GRB intrinsic flux would be reduced of a factor $\sim2.3$ ($\sim\num{1.5e6}$) at \SI{100}{\GeV} (\SI{1}{\TeV}). For this reason, the low energy threshold of MAGIC is crucial so that GRBs can be observed in an energy range where the intrinsic spectrum is not so strongly affected by the EBL. \\
The list of GRBs observed by MAGIC after the upgrade of the automatic procedure, between 2013 and April 2015, can be found in \cite{carosietal2015}. The GRBs observed from May 2015 to December 2018 instead are listed in Table\,\ref{tab:grb}. It can be noted that the follow-up of GRBs detected by the LAT has increased in the last years. Moreover, for many of these events the observation was performed even at late times. \\
The analysis of the GRBs listed in Table\,\ref{tab:grb} is currently ongoing using the standard MAGIC analysis package MARS. These GRBs, together with the ones observed between 2013 and April 2015, will be the topic of a forthcoming publication. Even in the case that no significant VHE signal will be detected, the flux upper limits can give important information. As an example, they can be used to reduce the phase space and the degeneracy between the several parameters of GRB emission models or they can be compared with specific models, like self-synchrotron Compton (SSC) by electrons.

\section{Conclusions and prospects}
Thanks to its remarkable performance and slewing capabilities, MAGIC is well suited to perform the follow-up of GRBs at VHE. In the last years, MAGIC has increased even more its effort on GRB searches, dedicating a considerable fraction of observation time. The upgrade of the automatic GRB procedure in 2013 made possible to reduce technical problems and the MAGIC GRB group is adapting the follow-up strategy, taking into account the increasing information coming from other experiments, especially \textit{Fermi}-LAT. In order to increase the chances of detection, several improvements are expected both on the analysis and hardware side. In particular, for the latter case, tests are planned to include the use of the so-called Sum-Trigger II into the automatic GRB procedure. The consequent lowered energy threshold, down to $\sim\SI{30}{\GeV}$, will give the possibility to observe sources even farther away, GRBs included, in an energy range not affected by the EBL. Moreover, it would increase the energy overlap between MAGIC and \textit{Fermi}, allowing for direct comparison and cross calibration of signals observed by different instruments.

\section{Acknowledgments}
The authors gratefully acknowledge financial support from the agencies and organizations listed here: \url{https://magic.mpp.mpg.de/acknowledgments_ICRC2019/}.

\end{document}